\documentclass[conference]{IEEEtran}

\usepackage{graphics, array, url, hyperref, epsfig, amsmath, amssymb, graphicx}
\usepackage{multirow}
\usepackage{subfigure}
\usepackage{graphicx,cite,color}

\begin{document}

\title{Private, Fair, and Verifiable Aggregate Statistics for Mobile Crowdsensing in Blockchain Era}


 \author{\IEEEauthorblockN{Miao He$^{\dagger}$, Jianbing Ni$^{\dagger}$, Dongxiao Liu$^{\ddagger}$, Haomiao Yang$^{\natural*}$, and Xuemin (Sherman) Shen$^{\ddagger}$\\}
   \IEEEauthorblockA{$^{\dagger}$Department of Electrical \& Computer Engineering, Queen's University, Kingston, Canada K7L 3N6 \\}
   \IEEEauthorblockA{$^{\ddagger}$ Department of Electrical \& Computer Engineering, University of Waterloo, Waterloo, Canada N2L 3G1\\}
 \IEEEauthorblockA{$^{\natural}$ University of Electronic Science and Technology of China, Chengdu, China 611731 \\}
 Email: \{19mh48, jn72\}@queensu.ca, \{dongxiao.liu, sshen\}@uwaterloo.ca, haomyang@uestc.edu.cn}

\maketitle

\begin{abstract}

In this paper, we propose FairCrowd, a private, fair, and veriﬁable framework for aggregate statistics in mobile crowdsensing based on the public blockchain. In specific, mobile users are incentivized to collect and share private data values (e.g., current locations) to fulﬁll a commonly interested task released by a customer, and the crowdsensing server computes aggregate statistics over the values of mobile users (e.g., the most popular location) for the customer. By utilizing the ElGamal encryption, the server learns nearly nothing about the private data or the statistical result. The correctness of aggregate statistics can be publicly veriﬁed by using a new efﬁcient and veriﬁable computation approach. Moreover, the fairness of incentive is guaranteed based on the public blockchain in the presence of greedy service provider, customers, and mobile users, who may launch payment-escaping, payment-reduction, free-riding, double-reporting, and Sybil attacks to corrupt reward distribution. Finally, FairCrowd is proved to achieve veriﬁable aggregate statistics with privacy preservation for mobile users. Extensive experiments are conducted to demonstrate the high efﬁciency of FairCrowd for aggregate statistics in mobile crowdsensing.

\end{abstract}

\section{Introduction} \label{sec1}
Mobile crowdsensing enables a group of individuals to collect and share telemetry data and other sensor readings using modern mobile devices, such as smart phones, vehicles, and wearable devices \cite{Capponi19}. With these data in hand, the crowdsensing server computes aggregate statistics to fulfill data-intensive tasks released by its customers \cite{Gibbs17}. By utilizing the intelligence of a crowd, mobile crowdsensing can significantly improve the quality of the collected data and the credibility of the statistical results, thereby enabling a broad range of applications \cite{Hui17}. For example, Google traffic collects real-time location information from mobile phones to identify traffic congestion and generate real-time traffic maps, which help drivers route along the uncrowded areas. Fitness tracking collects data on mobile phones to show their users' physical activities, so as to enable users to acquire and compare energy costs with average values. Walmart collects customers' purchases to learn the preferences for product recommendation and optimize products' exhibition on shelves to ease customers and increase sales.

The mobile crowdsensing service is interested in collecting the raw data from mobile users, and generates aggregate statistics for its customers' convenience, such a service often ends up with the privacy violation of mobile users \cite{NICM,Yang15}. The data values collected from the surrounding environments of mobile users pose a series of security and privacy risks: malicious attackers may expose the collected data; the service provider may abuse or sell the raw data for profit; and intelligent agencies may appreciate the data for knowledge discovery and massive surveillance. To migrate the risks on security corruption and privacy violation, mobile users may refuse to share their collected data in mobile crowdsensing.
To encourage their participation, monetary incentives are usually offered to assign rewards to the mobile users who make real efforts \cite{Wu19,Liu19}. Unfortunately, the service provider may fail to keep fair towards all the participants or resolve disputes. First of all, the customers may request data collection, but escape to pay the rewards they claimed (payment escaping), and the service is designed to bias on customers, such as Amazon M-Turk. Secondly, the service provider may also hide and possess part of rewards (payment-reduction), which is hard to detect in a fully distributed fashion \cite{Zheng17}. Thirdly, 
due to the reward temptation, mobile users may reap rewards without making contributions (free-riding), report duplicate data for repeated rewarding (double-reporting), or forge false identities for data sharing (Sybil). As the result, the fairness of reward distribution is corrupted.

To facilitate fair incentives, a straightforward solution is to employ an external trusted third party (TTP) to replace the service provider for reward host \cite{Zhang18}. However, finding a fully trusted entity in reality is difficult. Facebook's troubles and Snowden's revelations \cite{Snowden} have decreased human's trust on a single institution or government. It is desirable to reduce the reliance on a TTP in practice. The public blockchain \cite{Tschorsch16} is an open, distributed and transparent public ledger used to maintain a continuously growing list of transactions in cryptocurrency, e.g., Bitcoin and Ethereum. It is a chain of blocks and managed by multiple nodes in a peer-to-peer network. The blockchain offers decentralized transaction management and reliable transaction delivery in untrusted Internet \cite{Yu20,Su20}. Therefore, the blockchain is a potential solution to manage the rewards in a decentralized way. More importantly, the transactions in each block can be programmed to be executable codes, smart contract \cite{Steffen19}. The consensus protocol enforces automated execution of smart contracts, such that neither a single party nor a smart group of entities can interfere with the execution of a contract. In addition, blockchain naturally embodies a discrete notion of time \cite{Kosba16}, i.e., a clock, which increments whenever a new block is produced. The smart contracts and the trusted clock are crucial for attaining financial fairness in transactions and protocols, indicating that malicious contractual entities cannot prematurely abort from a protocol to refuse financial payment \cite{Lu18}.

In this paper, we propose FairCrowd, a blockchain-based framework for aggregate statistics in mobile crowdsensing that resolves the tension between privacy and fairness. A mobile user encrypts the data value before uploading it to the crowdsensing server, and the server performs aggregate statistics over the ciphertexts for the customer. The fairness is maintained in the presence of greedy mobile users, customers, and the service provider. Specifically, the contributions can be summarized in two folds.
\begin{itemize}
  \item We propose privacy-preserving and verifiable aggregate statistics, a type of secure computation on data values shared by different mobile users with correctness verification of statistical result by a designated verifier. The distinguished feature is that the statistical result verification and data authentication are achieved based on the homomorphic signature, simultaneously.
      The crowdsensing server learns nearly nothing about the data values or the statistical result, unless the mobile users or the customer discloses them. Moreover, to ensure the correctness of aggregate statistics in the presence of the untrusted crowdsensing server, the privately verifiable homomorphic signatures are generated by the mobile users and broadcasted to the network nodes on the blockchain. Thus, the customer is delegated to the capability of verifying the correctness of the statistical result without re-computing the result from the raw data by herself.

  \item We utilize smart contracts to maintain the fairness in the presence of greedy customers, mobile users, and service provider in mobile crowdsensing. Due to the permissionless access of blockchain and the automated execution of smart contract, the potential attacks can be prevented, including payment-escaping, payment-reduction, free-riding, double-reporting, and Sybil attacks.

\end{itemize}

The remainder of this paper is organized as follows. We present system and security models in section \ref{sec2}. We propose FairCrowd in section \ref{sec3}, and demonstrate its security features in section \ref{sec4}, followed by performance evaluation in section \ref{sec5}. Finally, we conclude our paper in section \ref{sec7}.

\section{System and Security Models} \label{sec2}
We present the system model and the security model of FairCrowd, and identify our design goals.
\subsection{Blockchain-based Mobile Crowdsensing}
Blockchain-based mobile crowdsensing consists of four entities, namely, a service provider, customers, mobile users and a public blockchain, as depicted in Fig. 1.

\begin{figure}
\centerline{\includegraphics[width=0.45\textwidth]{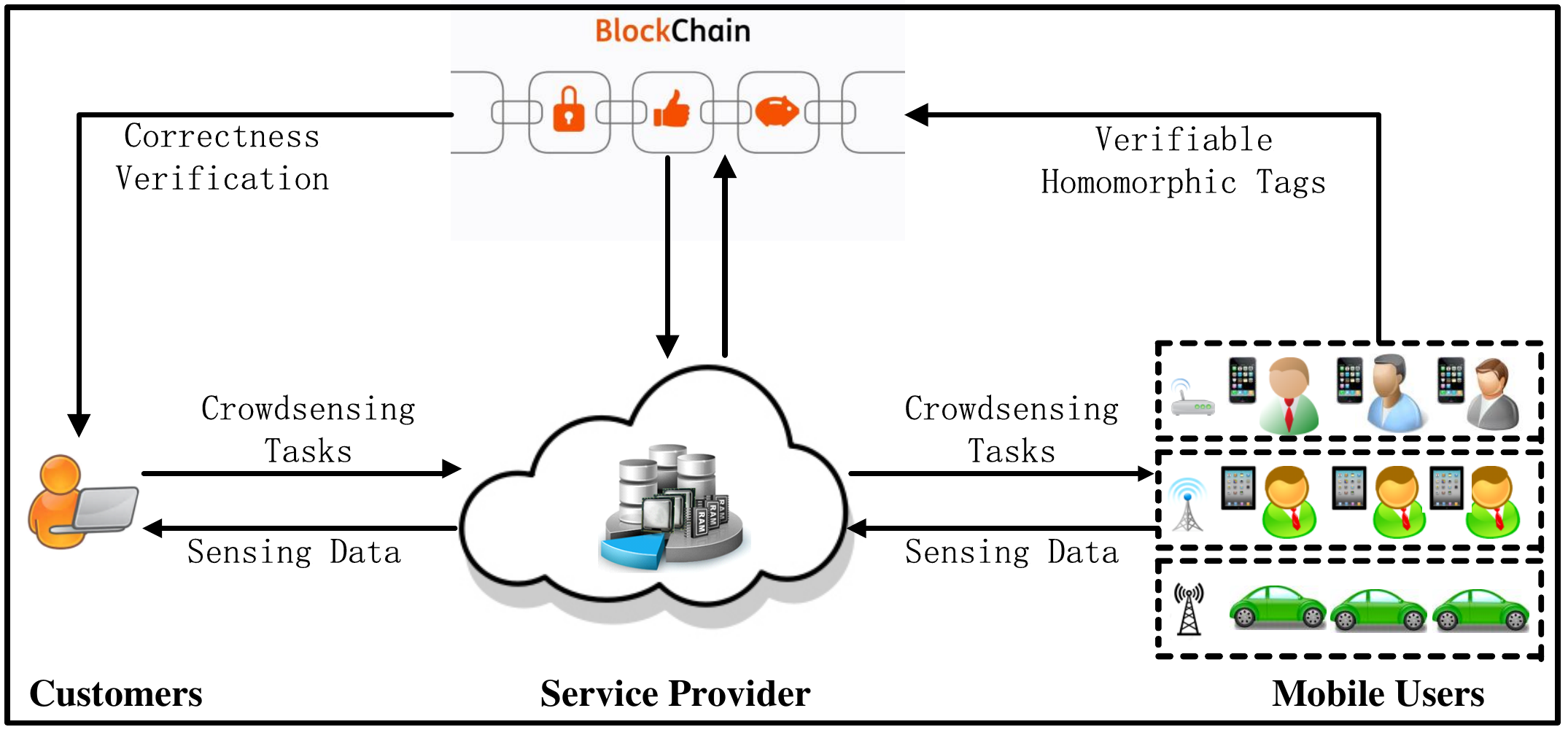}}
\caption{System Model.}
\label{fig:one}
\vspace{-0.2in}
\end{figure}

{Service Provider}: The service provider provides its customers with the mobile crowdsensing service.  It is responsible for releasing crowdsensing tasks for customers, recruiting mobile users for data collection based on their interests, aggregating sensing data values of mobile users, and finally distributing rewards to mobile users based on a pre-defined reward policy.

 {Customers}: The customers can be individuals, corporations, or organizations. They have some crowdsensing tasks to accomplish, such as real-time traffic monitoring, indoor floor plan reconstruction, and social recommendation, but they do not have sufficient capability to fulfill by themselves. The customers release their tasks on the crowdsensing server, provide incentives to reward mobile users for their contributions on data collection, and obtain the statistical results from the service provider.

{ Mobile Users}: Each mobile user has devices to perform crowdsensing tasks, e.g., smart phones, tablets, vehicles, laptops, and other items with sensors, computing units and storage spaces. The mobile users can participate in crowdsensing tasks by collecting data from environment, analyze data, and upload the sensing data values to the crowdsensing server.

{ Blockchain}: The blockchain is a public and decentralized ledger managed by the network nodes, i.e., miners \cite{Li17}. The network nodes verify all the validity of transactions and add the valid transactions into the block. The blockchain offers decentralized reward management for both customers and mobile users, and smart contracts are used to enforce automated payment of rewards.

\subsection{Security Model}
The service provider may attempt to steal rewards, hide rewards, manipulate reward assignment, or lie to the customers. The service provider is interested in the sensing data values and the statistical results. The customers concern their privacy leakage, and prefer to encrypt the crowdsensing tasks before releasing \cite{NITDSC18}. They are greedy that they would not honestly assign the rewards to mobile users based on the reward policy, instead, they may find various excuses to refuse to pay or deduce the rewards, such as prematurely aborts. Mobile users concern their privacy leakage and are greedy to the rewards. They may leverage a variety of attacks to reap rewards, such as free-riding, double-reporting, and Sybil attacks. In free-riding attacks, mobile users may reap rewards without making real efforts, such as replaying the sensing data generated by other mobile users; in double-reporting attacks, mobile users may submit the sensing data more than once to claim repeated rewards; and mobile users can fake identities to submit multiple copies of sensing data to obtain more rewards in Sybil attacks.

\subsection{Design Goals}
We attempt to realize privacy, fairness, and verifiability in FairCrowd on top of the existing architecture of public blockchain.

\begin{itemize}
  \item { Privacy}. The aggregate statistics in mobile crowdsensing shall be protected in case the inputs or the outputs are leaked to the unauthorized entities. The inputs, i.e., the sensing data values of mobile users, will be only shared with the customer. The output, i.e., the statistical result, computed by the crowdsensing server will be only known by the customer that pays the rewards to the participating mobile users.
  \item { Fairness}. The mobile crowdsensing service is fair, if (i) the customers cannot prematurely abort the protocols to escape from payment; (ii) the service provider cannot claim or leave the rewards to itself; (iii) the mobile users are unable to use free-riding, double-reporting or Sybil attacks to acquire more rewards than the amount they deserve to have.
  \item { Verifiability}. The aggregate statistics are verifiable if the customers are able to verify the correctness of statistical results, while ensuring that the sensing data are authenticated by mobile users. The verifiability of rewards refers to that the reward assignment is publicly verifiable.
\end{itemize}

\section{FairCrowd} \label{sec3}
In this section, we propose our FairCrowd, which consists of a private and verifiable aggregate statistical scheme and a smart contract.

\subsection{Linearly Aggregate Statistics with Verifiable Computation}
We first achieve linearly aggregate statistics with verifiable computation for a group of mobile users. This primitive is built atop the blockchain that acts as a bulletin board to maintain homomorphic signatures of mobile users. The homomorphic encryption scheme is leveraged to achieve privacy-preserving linear aggregation over individual private data $m_i=(m_{i1},m_{i2},\cdots,m_{il})$, and $m_{ij}$ is an independent data value of a dimension, for $j=1$ to $l$. We utilize the ElGamal encryption scheme as an example for data encryption, and the $\Sigma$-protocol to prove that the messages in the ciphertexts and signatures are identical. Formally, our private and verifiable aggregate statistical scheme (PVAS) is presented as follows.

\begin{itemize}
  \item { PVAS.ParGen}. Let $p$ be a large prime with $\lambda$ bits and ($\mathbb{G}_1,\mathbb{G}_2,\mathbb{G}_T$) be three cyclic groups of the order $p$. $\hat{e}: \mathbb{G}_1 \times \mathbb{G}_2 \rightarrow \mathbb{G}_T$ is the type-III bilinear pairing. $g,g_1,\cdots,g_l$ are generators of $\mathbb{G}_1$, and $h,h_1,\cdots,h_l$ are generators of $\mathbb{G}_2$. ${H}: \{0,1\}^* \rightarrow \mathbb{G}_1$ is a collision-resistant hash function.
  \item{ PVAS.KeyGen}. The customer randomly selects $a \in \mathbb{Z}_p$ as the secret key, and generates the corresponding public key $A=h^a$. Each mobile user randomly chooses $u_i \in \mathbb{Z}_p$ as the secret key, and computes the corresponding public key $U_i=h^{u_i}$. The crowdsensing server randomly selects $v \in \mathbb{Z}_p$ as the secret key, and computes the corresponding public key $\Lambda=h^v$.
  \end{itemize}

\begin{picture}(-10,608)
\put(0,10){\framebox(248,600)}
\put(0,554){\makebox(100,100)[l]{~~~~~~~~~~~~~~~~~~Smart Contract { CS-FairCrowd}}}
\put(0,542){\makebox(200,95)[l]{~~~~~\textbf{Init}: Set {state:=INIT}, {Task}$:=\{\}$, {AU}$:=\{\}$, {RU}$:=\{\}$,}}
\put(0,530){\makebox(200,95)[l]{~~~~~~~~~~~{{RUP}$:=\{\}$, { Param:=PVAS.ParGen}$(1^{\lambda})$}.}}
\put(0,518){\makebox(200,95)[l]{~\textbf{Create}: Upon receiving (``Create", $N$, $task$, $A$, Reward,}}
\put(0,506){\makebox(200,95)[l]{~~~~~~~~~~ $T_1,T_2,T_3,T_4$) from a customer $\mathcal{C}$:}}
\put(0,494){\makebox(200,95)[l]{~~~~~~~~~~~~~~~Assert {state=INT}.}}
\put(0,482){\makebox(200,95)[l]{~~~~~~~~~~~~~~~Assert current time $T\leq T_1$.}}
\put(0,470){\makebox(200,95)[l]{~~~~~~~~~~~~~~~Assert {ledger}$\mid\mathcal{C}\mid\geq$ \$Reward.}}
\put(0,458){\makebox(200,95)[l]{~~~~~~~~~~~~~~~{ledger} $\mid\mathcal{C}\mid$:={ledger}$\mid\mathcal{C}\mid$--\$Reward.}}
\put(0,446){\makebox(200,95)[l]{~~~~~~~~~~~~~~~Set {state:=CREATED}.}}
\put(0,434){\makebox(200,95)[l]{~~~~~~~~~~~~~~~Set {Accept}:=0.}}
\put(0,422){\makebox(200,95)[l]{~~~~~~~~~~~~~~~{Task}:={Task}$\cup$\{$\mathcal{C},N$,$A$,Reward,{Accept},$T_{j=1-4}$\}.}}
\put(0,410){\makebox(200,95)[l]{~\textbf{Accept}: Upon receiving (``Accept", $\mathcal{U}_i,N,R_i$) from a}}
\put(0,398){\makebox(200,95)[l]{~~~~~~~~~~ mobile user $\mathcal{U}_i$:}}
\put(0,386){\makebox(200,95)[l]{~~~~~~~~~~~~~~~Assert {state=CREATED}.}}
\put(0,374){\makebox(200,95)[l]{~~~~~~~~~~~~~~~Assert $T_1 \leq T \leq T_2$.}}
\put(0,362){\makebox(200,95)[l]{~~~~~~~~~~~~~~~Assert \$R$_i >$0.}}
\put(0,350){\makebox(200,95)[l]{~~~~~~~~~~~~~~~Assert {ledger}$\mid\mathcal{U}_i\mid\geq$ \$R$_i$.}}
\put(0,338){\makebox(200,95)[l]{~~~~~~~~~~~~~~~{ledger} $\mid\mathcal{U}_i\mid$:={ledger}$\mid\mathcal{U}_i\mid$--\$R$_i$.}}
\put(0,326){\makebox(200,95)[l]{~~~~~~~~~~~~~~~Set {Accept}:={Accept}+1.}}
\put(0,314){\makebox(200,95)[l]{~~~~~~~~~~~~~~~Set {state}$_i$:={ACCEPTED}.}}
\put(0,302){\makebox(200,95)[l]{~~~~~~~~~~~~~~~{AU}$_N$:={AU}$_N\cup\{\mathcal{U}_i$\}.}}
\put(0,290){\makebox(200,95)[l]{~~\textbf{Claim}: Current time $T=T_2$:}}
\put(0,278){\makebox(200,95)[l]{~~~~~~~~~~~~~~~Assert {state}$_i$={ACCEPTED}.}}
\put(0,266){\makebox(200,95)[l]{~~~~~~~~~~~~~~~Assert the fulfillment of the task $N$.}}
\put(0,254){\makebox(200,95)[l]{~~~~~~~~~~~~~~~Set {state:=CLAIMED}.}}
\put(0,242){\makebox(200,95)[l]{~\textbf{Upload}: Upon receiving (``Report", $\mathcal{U}_i,N,c_i,d_i,\sigma_i,e_i,rk_i,$}}
\put(0,230){\makebox(200,95)[l]{~~~~~~~~~~~~$\mathcal{PK}_i$) from $\mathcal{U}_i$:}}
\put(0,218){\makebox(200,95)[l]{~~~~~~~~~~~~~~~Assert {state=CLAIMED}.}}
\put(0,206){\makebox(200,95)[l]{~~~~~~~~~~~~~~~Assert $T_{2} \leq T\leq T_{3}$.}}
\put(0,194){\makebox(200,95)[l]{~~~~~~~~~~~~~~~Assert $\mathcal{U}_i\in${AU}$_N$.}}
\put(0,182){\makebox(200,95)[l]{~~~~~~~~~~~~~~~Assert $\mathcal{PK}_i=1$.}}
\put(0,170){\makebox(200,95)[l]{~~~~~~~~~~~~~~~Set {state}$_i$:={UPLOADED}.}}
\put(0,158){\makebox(200,95)[l]{~~~~~~~~~~~~~~~Set {ledger} $\mid\mathcal{U}_i\mid$:={ledger}$\mid\mathcal{U}_i\mid$+\$R$_i$.}}
\put(0,146){\makebox(200,95)[l]{~~~~~~~~~~~~~~~{RU}$_N$:={RU}$_N\cup\{\mathcal{U}_i$\}.}}
\put(0,134){\makebox(200,95)[l]{~~~~~~~~~~~~~~~{RUP}$_N$:={RUP}$_N\cup\{(\mathcal{U}_i,N,\sigma_i,e_i,rk_i)$\}.}}
\put(0,122){\makebox(200,95)[l]{~\textbf{Reward}: $T_{3} \leq T\leq T_{4}$ and {AU}$_N$={RU}$_N$:}}
\put(0,110){\makebox(200,95)[l]{~~~~~~~~~~~~~~~Set {state:=FULFILLED}.}}
\put(0,98){\makebox(200,95)[l]{~~~~~~~~~~~~~~~Set {ledger} $\mid\mathcal{U}_i\mid:=${ledger}$\mid\mathcal{U}_i\mid$+\$Reward$_i$.}}
\put(0,86){\makebox(200,95)[l]{~~~~~~~~~~~~~~~Assert \$Reward=$\sum_{i=1}^n$\$Reward$_i$.}}
\put(0,74){\makebox(200,95)[l]{~~~~~~~~~~~~~~~Set {state:=FINISHED}.}}
\put(0,62){\makebox(200,95)[l]{~\textbf{Penalty}: $T_{3} \leq T\leq T_{4}$ and {AU}$_N\supset${RU}$_N$:}}
\put(0,50){\makebox(200,95)[l]{~~~~~~~~~~~~~~~Set {state:=UNFULFILLED}.}}
\put(0,38){\makebox(200,95)[l]{~~~~~~~~~~~~~~~{ledger}$\mid\mathcal{U}_i\mid:=${ledger}$\mid\mathcal{U}_i\mid$+\$R$^*_i$, for $\mathcal{U}_i \in ${RU}$_N$.}}
\put(0,21){\makebox(200,95)[l]{~~~~~~~~~~~~~~~Assert $\sum\limits_{i\in\{{\texttt{AU}}_N-{\texttt{RU}}_N\}}$\$R$_i$ =$\sum\limits_{i\in\{{\texttt{RU}}_N\}}$\$R$^*_i$.}}
\put(0,6){\makebox(200,95)[l]{~~~~~~~~~~~~~~~Set {state:=ABORTED}.}}
\put(0,-6){\makebox(200,95)[l]{~~\textbf{Timer}: If {state=ABORTED} and $T>T_{4}$;}}
\put(0,-18){\makebox(200,95)[l]{~~~~~~~~~~~~~~~Set {ledger} $\mid\mathcal{C}\mid:=${ledger}$\mid\mathcal{C}\mid$+\$Reward.}}
\put(0,-30){\makebox(200,95)[l]{~~~~~~~~~~~~~~~Set {state:=ABORTED.}}}
\put(0,-10){\makebox(250,20){Alg. 1.  Smart Contract {CS-FairCrowd}}}
\end{picture}

\begin{itemize}
  \item{PVAS.SigEnc}. Each mobile user first encrypts the private data $m_i=(m_{i1},m_{i2},\cdots,m_{il})$ by randomly picking $r_{i1},r_{i2}, \cdots,r_{il}\in \mathbb{Z}_p$ to compute $c_i=(c_{i1},c_{i2},\cdots,c_{il})=(h_1^{m_{i1}}A^{r_{i1}},h_2^{m_{i2}}A^{r_{i2}},\cdots,h_l^{m_{il}}A^{r_{il}})$, and $d_{i}=(d_{i1},d_{i2},\cdots,d_{il})=(h^{r_{i1}},h^{r_{i2}},\cdots,h^{r_{il}})$. The user then chooses a random value $\tau_{i} \in \mathbb{Z}_p$ to compute $\sigma_i=(H(N||A)^{\tau_i}\prod_{j=1}^l g_j^{m_{ij}})^{u_i}$, $e_i=h^{\tau_i}$, and $rk_i=A^{u_i^{-1}}$, where $N$ is a random value as the task identifier in mobile crowdsensing. After that, the user generates the following zero-knowledge proof $\mathcal{PK}_i:\{(u_i,r_i,\tau_i,m_{i1},m_{i2},\cdots,m_{il}): c_{i1}=h_1^{m_{i1}}A^{r_i} \land \cdots \land c_{il}=h_l^{m_{il}}A^{r_i} \land d_{i1}=h^{r_{i1}} \land \cdots\land d_{il}=h^{r_{il}}  \land  \sigma_i=(H(N||A)^{\tau_i}\prod_{j=1}^l g_j^{m_{ij}})^{u_i} \land e_i=h^{\tau_i}\}$. Finally, the user sends $(c_i,d_i,rk_i)$ to the crowdsensing server, and broadcasts $(c_i,d_i,rk_i,\sigma_i,e_i,\mathcal{PK}_i)$ to the blockchain network. The nodes on the blockchain verify the validity of $\mathcal{PK}_i$ and insert $(\sigma_i,e_i)$ into the new block as a transaction.
  \item{PVAS.Agg}. Assume the linear function $f$ is parsed as $\{\omega_i\}_{1 \leq i \leq n}$. After receiving $n$ individual ciphertexts $((c_1,d_1),(c_2,d_2),\cdots,(c_n,d_n))$, the crowdsensing server aggregates all the received ciphertexts as $c_j=\prod_{i=1}^n c_{ij}^{\omega_i}$, and $d_j=\prod_{i=1}^n d_{ij}^{\omega_i}$, for $1 \leq j \leq l$. The crowdsensing server also aggregates the individual signatures as $\sigma=\prod_{i=1}^n\hat{e}(\sigma_i,rk_i)^{v\omega_i}$ and $e=\prod_{i=1}^n e_i^{v\omega_i}$.
  \item {PVAS.Dec}. For each $1 \leq j \leq l$, the customer uses $a$ to decrypt $h_j^{\sum_{i=1}^n m_{ij}\omega_i}=c_jd_j^{-a}$ and leverages the Pollard's lambda method to recover $m^*_j=\sum_{i=1}^n m_{ij}\omega_i$.
  \item{PVAS.Verify}. The customer checks whether $\sigma=\hat{e}(H(N||A),e)^a\hat{e}(\prod_{j=1}^l g_j^{m^*_{j}},\Lambda)^a$. If the equation holds, all $m^*_j$ are valid; otherwise, the customer rejects $m^*_j$ for $1 \leq j\leq l$.
\end{itemize}

\subsection{Smart Contract {CS-FairCrowd}}
We describe the proposed smart contract {CS-FairCrowd} to prevent the misbehavior of greedy customers, mobile users and service provider. {CS-FairCrowd} is given in Alg. 1.

\subsection{FairCrowd}
FairCrowd consists of four phases, namely, {Service Initialization}, {Task Releasing}, {Data Uploading}, and {User Rewarding}.

{Service Initialization}. The service provider bootstraps the whole mobile crowdsensing service by calling {PVAS.ParGen}. $H_1:\{0,1\}^*\rightarrow \mathbb{Z}_p$ is the collision-resistant hash function. The customer $\mathcal{C}$, a mobile user $\mathcal{U}_i$, and the crowdsensing server generate their secret-public key pairs using {PVAS.KeyGen}, separately. $\mathcal{C}$'s secret-public key pair is $({a,A})$, and $\mathcal{U}_i$'s secret-public key pair is $(u_i,U_i)$. The crowdsensing server's secret-public key pair is $(v,\Lambda)$.

{Task Releasing}. When $\mathcal{C}$ has a task $task$ to be crowdsourced to mobile users, $\mathcal{C}$ defines the reward policy and generates the task request $\mathcal{T}$=(``Create", $N$, $task$, $A$, Reward, $T_1,T_2,T_3,T_4$). $N$ is the task identifier. $task$ defines the detailed task goals, which include the sensing areas, time period, content, quantity, and other demands required to demonstrate. Reward denotes the reward policy and the number of rewards, \$Reward, to attract mobile users. $T_1,T_2,T_3,T_4$ are timeouts declared based on time property of the blockchain. {AU} stores the list of mobile users who have accepted the task. {RU} stores the list of mobile users who have reported the collected data, and {RUP} keeps the list of detailed information about the sensing data. Finally, $\mathcal{C}$ sends $\mathcal{T}$ to the crowdsensing server and broadcasts it to the blockchain network. The crowdsensing server maintains $\mathcal{T}$, and the network nodes perform {CS-FairCrowd.Init} and {CS-FairCrowd.Create} to create the releasing task.

{Data Uploading}. If a mobile user $\mathcal{U}_i$ is interested in the task $N$, $\mathcal{U}_i$ deposits the amount of coins $R_i$ as margins, indicating that $\mathcal{U}_i$ commits to make the uploading honestly. The amount can be determined by the mobile user or the crowdsensing server. $\mathcal{U}_i$ forwards an acceptance message $\mathbb{C}$=(``Accept", $\mathcal{U}_i, N, R_i$) to the crowdsensing server and the blockchain network. The network nodes perform {CS-FairCrowd.Accept} and {CS-FairCrowd.Claim}. The methods to estimate the task fulfillment are various, for example, the number of collected reports should reach a threshold, or there should be at least one mobile user in each sensing subarea. Here we consider a simple case that if all the accepted mobile users honestly upload their data, the crowdsensing task is fulfilled. The number of accepted mobile users is $n$.

If the task state is {CLAIMED}, $\mathcal{U}_i$ collects the data from the environment based on the demand in $task$ and generates the sensing data $m_i$. Before uploading $m_i$, $\mathcal{U}_i$ uses {PVAS.SigEnc} to compute the ciphertext $(c_i,d_i)$ of $m_i$, the re-sign key $rk_i$, the privately verifiable homomorphic signature $(\sigma_i,e_i)$, and the zero-knowledge proof $\mathcal{PK}_i$.
Then, $\mathcal{U}_i$ forwards $\mathcal{P}_i$=(``Report", $\mathcal{U}_i,N,$ $c_i,d_i,\sigma_i,e_i,rk_i,\mathcal{PK}_i$) to the crowdsensing server and the blockchain network. The network nodes call {CS-FairCrowd.Upload} and the crowdsensing server calls {PVAS.Agg} to generate $\mathcal{P}=(c_j,d_j,\sigma,e)$. Finally, the customer $\mathcal{C}$ executes {PVAS.Dec} to obtain $m^*_j=\prod_{i=1}^n e_i^{\omega_i}$, for $1 \leq j \leq l$. Also, the correctness of statistical result $m^*_j$, for $1 \leq j \leq l$, can be verified by using {PVAS.Verify}.

\section{Security Analysis}  \label{sec4}
In this section, we discuss the desirable properties of FairCrowd, i.e., privacy, fairness, and verifiability.

{Privacy}. The ElGamal encryption is leveraged to encrypt the private data before uploading. Semantic security can be reached as long as the decisional Diffie-Hellman (DDH) assumption holds \cite{Tsiounis98}. Also, the statistical result is still encrypted by the ElGamal encryption. Therefore, if the DDH assumption holds, the collected data and the statistical result do disclose nothing about the privacy of mobile users.

In addition, the private data in the homomorphic signature $(\sigma_i,e_i)$ and the statistical result in the aggregated signature $(\sigma,e)$ are protected against the off-line guessing attacks. Specifically, $\tau_i$ is randomly picked to randomize $\sigma_i$, it is impossible for an adversary $\mathcal{A}$ to test the verification equation of $(\sigma_i,e_i)$ without $\tau_i$. Also, the verification of $(\sigma,e)$ needs the knowledge of $a$, the secret key of $\mathcal{C}$. Thereby, only $\mathcal{C}$ can verify the correctness of the statistical result.

{Fairness}. The misbehavior of fairness corruption in reward distribution, including payment escaping, payment-reduction, free-riding, double-reporting, and Sybil attacks, is prevented by {CS-FairCrowd}. The reasons are illustrated as follows:
\begin{itemize}
  \item {Payment escaping}. $\mathcal{C}$ commits payment and deposits rewards on the blockchain using {CS-FairCrowd}. The deposited rewards will be automatically transferred to the mobile users. $\mathcal{C}$ cannot interrupt the transactions once the rewards are deposited.

  \item {Payment-reduction}. The reward distribution is hosted by {CS-FairCrowd}, which distributes the committed rewards to the proper entities based on the policy. The service provider cannot illegally possess the rewards.

  \item {Free-riding}. The contributions of mobile users are recorded on the blockchain. If the mobile users refuse to upload their data, they will lose the deposited coins. Thus, if $\mathcal{U}_i$ does not make any contribution on data collection, no reward will be assigned to $\mathcal{U}_i$.

  \item {Double-reporting}. The double-reporting can be detected by both $\mathcal{C}$ and the service provider, since all the signatures are maintained on the blockchain. If a mobile user $\mathcal{U}_i$ uploads more than one report, the server can identify more than one signature of $\mathcal{U}_i$ on the blockchain.

  \item {Sybil}. The mobile users may adaptively update public keys to share the collected data. To prevent Sybil attacks, the operations of mobile users are separated into two steps, {CS-FairCrowd.Accept} and {CS-FairCrowd.Upload}. The public key that $\mathcal{U}_i$ uses in {CS-FairCrowd.Upload} should be identical to that in {CS-FairCrowd.Accept}. $\mathcal{U}_i$ cannot upload multiple data copes using different public keys, since the old public key cannot be recovered once it is updated to a new one. Thus, a mobile user can only use the same public key to accept the task and upload the private data. In doing so, the Sybil attack can be avoided. Also, to prevent $\mathcal{U}_i$ from maliciously accepting the task without sharing data, $\mathcal{U}_i$ needs to deposit the coins in {CS-FairCrowd.Accept}.

\end{itemize}

{Verifiability}. The correctness of both aggregate statistics and reward distribution are verifiable. The verification of reward distribution is guaranteed based on the blockchain, as all the transactions are transparent and publicly verifiable. The correctness verification of aggregate statistics is realized based on the extended homomorphic signatures that should satisfy the notion of unforgeability. The extended homomorphic signature extends the homomorphic signature that enables the homomorphic operations on the signatures with the proxy re-signing technique. In PVAS, due to the proxy re-signing, the unforgeability should be guaranteed in two levels, i.e., the unforgeability of homomorphic signatures $(\sigma_i,e_i)$ and the unforgeability of aggregated signature $(\sigma,e)$. Firstly, $(\sigma_i,e_i)$ is calculated by leveraging the BLS signature, whose unforgeability is reduced to the Computational Diffie-Hellman (CDH) problem in a Gap Diffie-Hellman (GDH) group \cite{Boneh2001}. Secondly, the unforgeability of $(\sigma,e)$ depends on the CDH problem in $\mathbb{G}_T$ \cite{Bao2003}, i.e., given $g \in \mathbb{G}_1, h, h^{a}, h^{v}\in \mathbb{G}_2$, where $a,v \in \mathbb{Z}_p^*$, to compute $\hat{e}(g,h)^{av} \in \mathbb{G}_T$. If a probabilistic polynomial-time adversary $\mathcal{A}$ can break the unforgeability of $(\sigma,e)$ with a non-negligible advantage, there is an algorithm $\mathcal{B}$ to solve the CDH problem in $\mathbb{G}_T$.

Given $g \in \mathbb{G}_1, h, h^{a}, h^{v}\in \mathbb{G}_2$, where $a,v \in \mathbb{Z}_p^*$, the goal is to compute $\hat{e}(g,g)^{av} \in \mathbb{G}_T$. $\mathcal{B}$ can access the signing oracle $\mathcal{SO}$ which outputs the homomorphic signatures of mobile users, and interact with $\mathcal{A}$ as follows.

 \begin{itemize}
   \item $\mathcal{B}$ randomly chooses $u_i \in \mathbb{Z}_p^*$ to set the public key $U_i$ to $h^{u_i}$ and the re-sign key $rk_i$ to $h^{au_i}$. Then, $\mathcal{B}$ picks random $\gamma_j \in \mathbb{Z}_p^*$ to set the parameter $g_j$ to $g^{\gamma_j}$, for $1 \leq j\leq l$. Finally, $\mathcal{B}$ sends $(h_i,rk_i,g_1,\cdots, g_l)$ to $\mathcal{A}$.

   \item $\mathcal{A}$ queries the hash oracle to the hash of $(N,A)$. $\mathcal{B}$ randomly picks $\gamma \in \mathbb{Z}_p^*$ and returns $g^{\gamma}$ to $\mathcal{A}$.

   \item $\mathcal{A}$ queries $\mathcal{B}$ the homomorphic signatures under any public key $h^{u_i}$. $\mathcal{B}$ issues a signing query to $\mathcal{SO}$, randomly chooses $\tau_i\in \mathbb{Z}_p^*$, and returns $(\sigma_i,e_i)$ to $\mathcal{A}$.

   \item Finally, $\mathcal{A}$ produces a valid aggregated signature $(\bar{\sigma},\bar{e})$ on $\bar{m}^*_{j}$ that satisfies

       \begin{center}
        $\bar{\sigma}=\hat{e}(H(N||A),\bar{e})^a\hat{e}(\prod_{j=1}^l g_j^{\bar{m}^*_{j}},\Lambda)^a.$
       \end{center}

       The expected signature obtained from the honest signers is $(\sigma, e)$ on $m^*_{j}$. $(\sigma,e)$ also satisfies

       \begin{center}
        ${\sigma}=\hat{e}(H(N||A),{e})^a\hat{e}(\prod_{j=1}^l g_j^{{m}^*_{j}},\Lambda)^a.$
       \end{center}

   \begin{itemize}
   \item If $\bar{m}^*_{j}\neq {m}^*_{j}$ for $1 \leq j \leq l$. We define that $\Delta m^*_{j}=\bar{m}^*_{j}-{m}^*_{j}$ for $1 \leq j \leq l$, it is the case that at least one of $\Delta m^*_{j}$ is nonzero. We divide the verification equation for $\bar{\sigma}$ by the equation for $\sigma$ and obtain

         \begin{center}
         $\bar{\sigma}/\sigma=\hat{e}(H(N||A),\bar{e}/e)^a\hat{e}(\prod_{j=1}^l g_j^{\bar{m}^*_{j}-{m}^*_{j}},\Lambda)^a.$
         \end{center}

       Since $H(N||A)=g^{\gamma}$, $e=h^{v\tau_i\omega_i}$ and $g_j=g^{\gamma_j}$ for $1 \leq j\leq l$, we have

          \begin{center}
          $\bar{\sigma}/\sigma=\hat{e}(g^{\gamma},h^{v\omega_i (\bar{\tau}_i-\tau_i)})^a\hat{e}(\prod_{j=1}^l g^{\gamma_j(\bar{m}^*_{j}-{m}^*_{j})},h^v)^a.$
          \end{center}

          Rearranging the equation yields

          \begin{center}
          $\hat{e}(g,h)^{av}=({\sigma}/\bar{\sigma})^{\gamma\omega_i (\bar{\tau}_i-\tau_i)+\sum_{j=1}^l \gamma_j(\bar{m}^*_{j}-{m}^*_{j})},$
         \end{center}

         which is the solution of the CDH problem.

    \item    Otherwise, $\bar{\sigma}/\sigma=\hat{e}(H(N||A),\bar{e}/e)^a$ and that

             \begin{center}
              $\hat{e}(g,h)^{av}=({\sigma}/\bar{\sigma})^{\gamma\omega_i (\bar{\tau}_i-\tau_i)},$
             \end{center}

               So we can solve the CDH problem.
     \end{itemize}
 \end{itemize}

\section{Performance Evaluation}  \label{sec5}
We evaluate the computational overhead of the crowdsensing server, mobile users, and customers, and the storage cost of the network nodes in FairCrowd.

We run microbenchmark on a HUAWEI MT2-L01 smartphone with Kirin 910 CPU and 1250M memory. The operation system is Android 4.2.2 and the toolset is Android NDK r8d. The smartphone is used to simulate the operations of mobile users and customers. A Thinkpad X1 Yoga laptop is setup to be the crowdsensing server with Intel Core i5-8265U CPU@1.8GHz and 16GB RAM, running the 64-bit Windows 10. The GNU Multiprecision Library and the Pairing-Based Cryptography Library are utilized to implement the cryptographic primitives. We build the polynomial ring $\mathbb{F}_p[x]$ with a 256-bit prime $p$ and use the Type III bilinear pairing and Barreto--Naehrig curve. SHA256 is utilized for hash function.

We release the task to collect air pollutant concentrations (i.e., PM2.5) of the cities in Ontario, Canada, and find 40 volunteers to upload the data based on the posted information  (http://www.airqualityontario.com/history/summary.php) using their mobile phones. The average reading of PM2.5 in Ontario is calculated for the customer. We collect the computation time of the corresponding entities processing these 40 data reports in FairCrowd, including the customer, the crowdsensing server, and the mobile users, as shown in Table I. The time cost of the mobile user in Table I is the time usage to perform data uploading on the HUAWEI smartphone, while the time cost of the crowdsensing server is the time usage to deal with 40 reports from mobile users.
In addition, with the increasing number of mobile users, the time cost of the server increases, while the time cost of the customer is constant. The scalability of FairCrowd is examined in Fig. 2.

\begin{table}
\caption{Computational Overhead of FairCrowd (Unit: milliseconds)}
\vspace{-0.1in}
\centering
\begin{tabular}{|c|c|c|c|c|}
\hline
\multirow{2}{*}{FairCrowd} & {Service } & {Task } &{Data } & {User}  \\

 & Initialization&Releasing & Uploading  & Rewarding  \\
\hline
Mobile User & 8&--&198&53\\
\hline
 Server & 1230&25&1641&--\\
\hline
Customer & 6 &84& 147&58\\
\hline
\end{tabular}
\end{table}

\begin{figure}
\centerline{\includegraphics[width=0.35\textwidth]{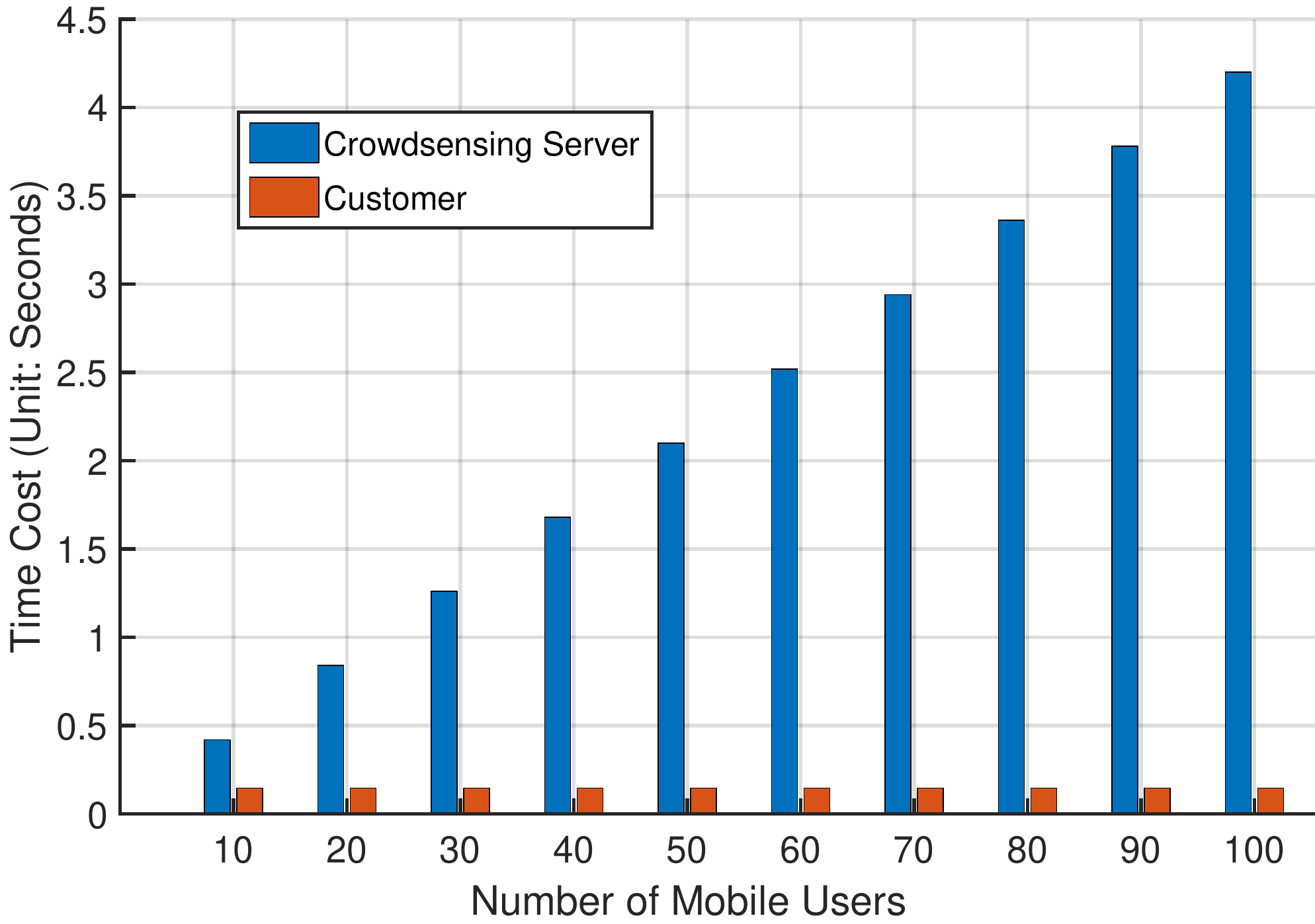}}
\caption{Scalability of FairCrowd.}
\label{fig:one}
\vspace{-0.2in}
\end{figure}

Also, we evaluate the storage overhead of the blockchain nodes and the crowdsensing server, which consists of on-chain storage and off-chain storage. In {CS-FairCrowd}, each blockchain node is required to maintain the task, the identities of mobile users, and state information about the task, once the smart contract is created. In {CS-FairCrowd.Create}, the blockchain node keeps about 128-byte task information. The data in {CS-FairCrowd.Upload} possesses nearly 390$n$ bytes space on each blockchain node for each task, where $n$ is the number of mobile users who accept the task. The off-chain storage is maintained on the crowdsensing server, which significantly reduces the on-chain storage overhead, and the zero-knowledge proof is utilized to keep the consistency of the data on the crowdsensing server and the blockchain nodes. The server needs to keep the encrypted private data of mobile users and the zero-knowledge proof $\mathcal{PK}_i$, the binary length of which is 288$ln$+192$n$ bytes, where $l$ is data dimensions.

\section{Conclusion}  \label{sec7}
In this paper, we have proposed a private, fair, and veriﬁable framework for aggregate statistics in mobile crowdsensing based on the blockchain. Aggregate statistics over the private data are enabled with efﬁcient correctness veriﬁcation of the statistical results. The fairness of mobile users are guaranteed to encourage them to participate in crowdsensing tasks, the success of which depends on the participation of honest mobile users. A new smart contract is designed to enforce fair reward distribution. We have demonstrated that FairCrowd achieves the properties of privacy, fairness, and veriﬁability, and is highly efﬁcient for aggregate statistics in mobile crowdsensing. For our future work, we will design an efﬁcient and fair data trading framework for a group of customers in mobile crowdsensing.

\vfill\eject

\end{document}